\documentclass[prd,aps,nofootinbib,floatfix,11pt]{revtex4}
\usepackage{amsmath,graphicx,epsfig,amssymb,dsfont,mathtools,float}
\usepackage[usenames]{color}
\usepackage{ulem}
\usepackage{bigstrut}
\usepackage{slashed}
\usepackage{multirow}
\usepackage{subfigure}
\usepackage{graphicx}
\usepackage[dvipdfmx,colorlinks=true,citecolor=blue,linkcolor=blue,urlcolor=blue]{hyperref}
\allowdisplaybreaks

\begin{document}

\title{Semi-invisible Hyperon Decays in the  Effective Lagrangian Approach}

\author{Lai Jiang}
\affiliation{School of Materials Science and Physics, China University of Mining and Technology, Xuzhou 221000, China}

\author{Ye Xing}
\email{Corresponding author. xingye\_guang@cumt.edu.cn}
\affiliation{School of Materials Science and Physics, China University of Mining and Technology, Xuzhou 221000, China}

\author{Yu Zhou}
\affiliation{School of Materials Science and Physics, China University of Mining and Technology, Xuzhou 221000, China}

\author{Xiao-hui Hu}
\email{Corresponding author. huxiaohui@cumt.edu.cn}
\affiliation{School of Materials Science and Physics, China University of Mining and Technology, Xuzhou 221000, China}

\begin{abstract}
We systematically investigate the semi-invisible decays of hyperons (hyperon $\to \pi(/\gamma)\ +$ invisible($\psi$)) in the Mesogenesis mechanism by the effective Lagrangian approach. The one-loop hadronic contribution of triangle diagrams with final-state interactions is fully examined in the present work.  
Our analysis indicates that the triangle diagram yield sizable corrections to the branching ratio that are as significant as those from tree diagrams. Especially for the $\Sigma^-\to \pi^-\psi$ and $\Xi^0 \to \pi^0 \psi$, their loop contributions cannot be ignored. Consequently, the branching ratios of hyperon hadronic semi-invisible decays are found to be of order $10^{-5}$, particularly for $\Sigma^+\to\pi^+\psi$, $\Xi^0\to\pi^0\psi$, and $\Xi^-\to\pi^-\psi$, whereas those of radiative semi-invisible decays are less than $10^{-7}$.
\end{abstract}
\maketitle

\section{Introduction}
Recently, low-scale baryogenesis mechanisms, referred to as Mesogenesis, have attracted increasing attention~\cite{Elor:2018twp,Alonso-Alvarez:2021qfd}. Such scenarios describe the interactions between dark baryonic sectors and the Standard Model (SM) particles, introducing a process in which SM hadrons produce dark matter baryons carrying a hidden baryon number, thereby simultaneously generating the baryon asymmetry and dark matter. What is more fascinating is that the typical scale of Mesogenesis mechanism is below electroweak scale, thus being directly probeable at the LHC, Belle, and BESIII experiments. The search for B-Mesogenesis by BABAR collaboration yield upper limits reaching $\mathcal{O}(10^{-6})$ for $B\to \Lambda \psi$ and $B^+\to p \psi$ in the range of $1.0<m_{\psi}<4.2$ GeV~\cite{BaBar:2023rer,BaBar:2023dtq}, while BELLE collaboration present a upper limits on the branching fractions in the range $\mathcal{O}(10^{-5})$ for the invisible decays of $B\to h \psi$, where $h=\pi, K, D_s, p$~\cite{Belle:2021gmc,Belle-II:2026tyb}. The BESIII collaboration has measured the invisible decays of hyperons including $\Lambda$, and $\Xi$ and provided respective upper limits~\cite{BESIII:2021slv,BESIII:2025sfl} reaching $\mathcal{O}(10^{-5})$. These experimental results can provide nontrivial constraints on the model parameters of such dark matter.

On the theoretical side~\cite{Alonso-Alvarez:2021oaj,Davoudiasl:2010am}, the effective four-fermion interactions emerging in the Mesogenesis models generate decays of the SM quark into a light diquark and a dark baryon $\psi$.  Accordingly, recent studies have paid considerable attention to semi-invisible decays of the b-quark. In Ref.~\cite{Khodjamirian:2022vta,Boushmelev:2023huu,Elor:2022jxy,Shi:2024uqs,Shi:2023riy}, the authors computed semi-invisible decays of B-mesons using the light-cone sum rule (LCSR) approach combined with the heavy-quark expansion (HQE) and perturbative QCD methods. Meanwhile, Ref.~\cite{He:2024iju,Zheng:2024tkj,Xing:2025pfw} investigated semi-invisible decays of b-baryons. However, the understanding of Mesogenesis processes involving light baryons remains incomplete. A recent development by Ref.~\cite{Alonso-Alvarez:2021oaj} established an effective Lagrangian for light baryons in the chiral representation, which providing a framework for investigating light-baryon Mesogenesis. At tree-level hadronic order, the authors placed constraints on the Wilson coefficients and branching fractions for semi-invisible decays. Nevertheless, the one-loop hadronic triangle diagram has only one or two more strong coupling vertices than the tree-level invisible decay diagram, and the strong coupling among hadrons is substantially larger than that between hadrons and dark baryons. Consequently, the one-loop hadronic scattering effects cannot be neglected and should be included in the analysis. To address this, the present work aims to compute semi-invisible decays of hyperons based on the effective Lagrangian approach~\cite{Cheng:2004ru,Cheng:2015naa,Li:2025wod} at both tree level and one-loop hadronic order.
Using the most recent experimental results, we will constrain the relevant Wilson coefficients and further evaluate the corresponding branching fractions.

The paper is organized as follow, in section~\ref{sec:ihd}, we introduce the semi-invisible decays of hyperons from phenomenological flavor symmetry. Next in section~\ref{sec:EFT}, we analyze the tree-level and one-loop hadronic diagrams relevant to semi-invisible decays of hyperons, based on the leading-order chiral effective Lagrangian. We discuss the Wilson coefficient and branching ratios in section~\ref{sec:analysis}. Finally, we conclude in the last section~\ref{sec:conclusion}.
\section{Semi-invisible decay processes: phenomenological analysis}
\label{sec:ihd}
\begin{figure}[H]
	\centering
	{\includegraphics[width=1\textwidth]{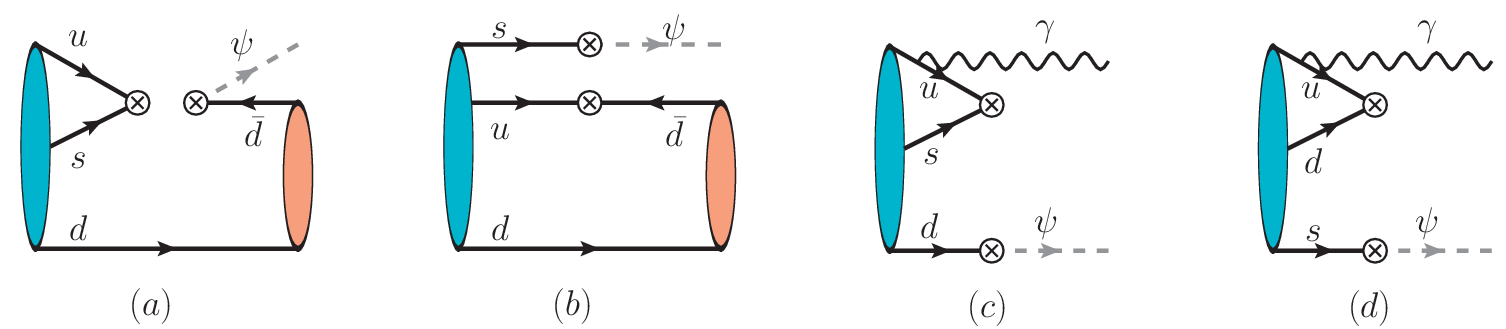}}
	\caption{The possible Feynman diagrams for the semi-invisible hyperon decays in quark-level, including meson or photon in final states. The circled cross symbol represents the effective interaction vertex between quarks and dark baryon.}
	\label{fig:dark}
\end{figure}

According to ~\cite{Alonso-Alvarez:2021oaj}, the decay of a light baryon into a light meson and dark baryon $\psi$ is generated by an exchange of heavy colour-triplet scalar field $\Psi$($\Phi$) or vector field $X_{\mu}$. The heavy mediators can be integrated out at the lower energies relevant for hadron decays, which leads to the local effective Lagrangian,
\begin{eqnarray}
&\mathcal{L}_{\rm{eff}} =C_{ab,c} \epsilon_{ijk}(u_{Ra}^i d_{Rb}^j)(\psi_R d_{Rc}^k)+C_{ab,c}^{\prime} \epsilon_{ijk}\epsilon_{\alpha\beta}(q_{La}^{i\alpha}q_{Lb}^{j\beta})(\psi_R d_{Rc}^k).
\end{eqnarray}
here $C^{\prime}_{ab,c}$ are the Wilson coefficients. $u$, $d$, $q$ represent two-component spinors in the Wely representation. $i$, $j$, $k$ are SU(3) color indices; $\alpha$, $\beta$ are SU(2) indices; $a$, $b$, $c$ refer to generation numbers, and $L$ and $R$ refer to left- and right-handed fields. We force on the interaction of light quarks, and factorizing out the external $\psi$ field, then the effective Lagrangian can be rewritten as,
\begin{eqnarray}\label{eq:Lagrangian_eff}
&\mathcal{L}_{\rm{eff}} =C_{ud_a,d_b}^R \mathcal{O}_{ud_a,d_b}^R \psi_R + C_{ud_a,d_b}^{L} \mathcal{O}_{ab,c}^{L} \psi_R,
\end{eqnarray}
where the effective coefficients $C^{L/R}$ are model-dependent, in the case of a vector mediator $X_{\mu}$, only the $C^L$ contribution is present, whereas for the scalar mediators, both $C^L$ and $C^R$ may contribute. $\mathcal{O}_{ab,c}^{L/R}$ is the local three-quark operator.
\begin{eqnarray}
\mathcal{O}_{ud_a,d_b}^R=\epsilon_{ijk}(u_{R}^i d_{Ra}^j)d_{Rb}^k,\ \mathcal{O}_{ud_a,d_b}^L=\epsilon_{ijk}(u_{L}^i d_{La}^j)d_{Rb}^k.
\end{eqnarray}
The good diquark ($u^id^j$) enforces an antisymmetric flavor wavefunction, which leads to the three-quark operator forming only a flavor octet representation $\mathcal{O}_8$. Combined with the possible transitions in Fig.~\ref{fig:dark}, we write the non-zero elements without chirality $(\mathcal{O}_8)^3_2=-C_{us,s}$, $(\mathcal{O}_8)^2_3=C_{ud,d}$, $(\mathcal{O}_8)^2_2=C_{ud,s}-C_{us,d}$, $(\mathcal{O}_8)^3_3=C_{ud,s}-C_{us,d}$. In flavor space, for the decays processes of light baryon ($\mathcal{B}$) to (light meson $\mathcal{P}$ $+$ dark baryon $\psi$) and to (photon $\gamma$ $+$ dark baryon), the flavor-invariant Hamiltonian~\cite{Savage:1989ub,He:1998rq,Shi:2017dto,Xing:2018bqt,Khlopov:1978id} reads directly:
\begin{align}
	\mathcal H_{\mathrm{eff}}^{\rm{SU(3)}}=a_1\mathcal{B}^i_j(\mathcal{O}_8)^j_k\mathcal{P}^k_i \psi+a_2\mathcal{B}^i_k(\mathcal{O}_8)^j_i\mathcal{P}^k_j \psi
+b_1 \mathcal{B}^i_j(\mathcal{O}_8)^j_i \psi \gamma,
\end{align}
here $a_i$ and $b_i$ are the non-perturbative coefficients, they correspond to hadronic semi-invisible decays and radiative semi-invisible decays of hyperons. Expanding the Hamiltonian constructed above, we collect the possible decay channels and their decay amplitudes in Tab.~\ref{tab:channels}. We consider only kinematically allowed decay processes and exclude the possibility of proton decay. All the amplitudes correspond to the Feynman diagrams. For instance, for processes $\Lambda^0\to \pi^0 \psi$, $\Sigma^0\to \pi^0 \psi$ and $\Sigma^+\to \pi^+\psi$, the amplitudes $C_{us,d}$ and $C_{ud,s}$ represent the annihilation diagram~\ref{fig:dark}.(a) and bosonic exchange diagram~\ref{fig:dark}.(b), respectively. Notably, the above processes must satisfy angular momentum conservation, in particular, in $\Lambda$ and $\Sigma^0$, the $ud$ pair carries definite spins of $0$ and $1$, respectively. Therefore, for  scalar $\Phi/\Psi$ or vector $X_{\mu}$ mediators, angular momentum conservation implies that the decay amplitudes of $\Lambda$ and $\Sigma^0$ are,
\begin{eqnarray}
&\mathcal{M}(\Lambda\xrightarrow{\Phi/\Psi}  \psi\gamma)=\frac{b_1(C_{us,d}^{L/R}-C_{ud,s}^{L/R})}{\sqrt6}, &\mathcal{M}(\Lambda\xrightarrow{X_{\mu}}  \psi\gamma)=\frac{b_1 C_{us,d}^L}{\sqrt6},\\
&\mathcal{M}(\Sigma^0\xrightarrow{\Phi/\Psi}  \psi\gamma)=\frac{b_1 C_{us,d}^{L/R}}{\sqrt2}, &\mathcal{M}(\Sigma^0\xrightarrow{X_{\mu}}  \psi\gamma)=\frac{b_1(C_{us,d}^L-C_{ud,s}^L)}{\sqrt2}.
\end{eqnarray}


When ignoring the phase-space effect, the relations between different decay channels can be reduced as
\begin{eqnarray}\label{eq:relations}
&&\Gamma\left(\Xi^-\to \pi^-\psi\right)=2\Gamma\left(\Xi^0\to \pi^0\psi\right), 3\Gamma(\Lambda^0\to \pi^0\psi)=\Gamma(\Sigma^0\to \pi^0\psi), \notag\\
&&\frac{\Gamma(\Sigma^-\to \pi^- \psi)}{2\Gamma(\Xi^0\to \pi^0 \psi)}=\Big|\frac{C_{us,d}^{L/R}-C_{ud,s}^{L/R}}{C_{us,s}^{L/R}}\Big|^2,
\frac{3\Gamma(\Sigma^0\xrightarrow{X_{\mu}}\psi\gamma)}{\Gamma(\Lambda^0\xrightarrow{X_{\mu}}\psi\gamma)}=\Big|\frac{C_{ud,s}^{L}-C_{us,d}^{L}}{C_{us,d}^{L}}\Big|^2,\notag\\
&&
\frac{6\Gamma(\Lambda^0\xrightarrow{X_{\mu}}\psi\gamma)}{\Gamma(\Xi^0\xrightarrow{X_{\mu}}\psi\gamma)}=\Big|\frac{C_{us,d}^{L}}{C_{us,s}^{L}}\Big|^2,
\frac{\Gamma(n\to \psi\gamma)}{\Gamma(\Xi^0\to \psi\gamma)}=\Big|\frac{C_{ud,d}^{L/R}}{C_{us,s}^{L/R}}\Big|^2.
\end{eqnarray}
In addition, the decay of $\Sigma^0$ is dominated by the electromagnetic decay $\Sigma^0\to \Lambda \gamma$, consequently, all other decay modes are highly suppressed. Therefore, the corresponding relation in Eq.~\ref{eq:relations} will be significantly broken.
\begin{table}
\caption{The possible and forbidden decay channels of light baryons into light mesons and dark baryon $\psi$.}
\label{tab:channels}
\centering
\begin{tabular}{cc|cc}
\hline
possible channel & amplitude &forbidden channel & amplitude \\ \hline
$\Lambda^0\to \pi^0 \psi$ & $\frac{(a_1+a_2)(C_{{us,d}}-C_{{ud,s}})}{2\sqrt{3}}$ & $\Lambda^0\to K^0 \psi$ & $\frac{(-2a_1+a_2)C_{{us,s}}}{\sqrt{6}}$ \\
$\Sigma^+\to \pi^+ \psi$ & $a_2(C_{{ud,s}}-C_{{us,d}})$ & $\Lambda^0\to \overline K^0 \psi$ & $-\frac{(a_1-2a_2)C_{{ud,d}}}{\sqrt{6}}$ \\
$\Sigma^0\to \pi^0 \psi$ & $\frac{(a_1+a_2)(C_{{ud,s}}-C_{{us,d}})}{2}$ & $\Sigma^+\to K^+ \psi$ & $a_2C_{{us,s}}$ \\
$\Sigma^-\to \pi^- \psi$ & $a_1(C_{{ud,s}}-C_{{us,d}})$ & $\Sigma^0\to K^0 \psi$ & $-\frac{a_2 C_{{us,s}}}{\sqrt{2}}$ \\
$\Xi^0\to \pi^0 \psi$ & $\frac{-a_1 C_{{us,s}}}{\sqrt{2}}$ & $\Sigma^0\to \overline K^0 \psi$ & $\frac{a_1C_{{ud,d}}}{\sqrt{2}}$ \\
$\Xi^-\to \pi^- \psi$ & $a_1C_{{us,s}}$ & $\Sigma^-\to K^- \psi$ & $-a_1 C_{{ud,d}}$ \\
$\Lambda^0 \to \psi \gamma$ & $\frac{b_1 (C_{{us,d}}-C_{{ud,s}})}{\sqrt{6}}$ & $\Xi^-\to K^- \psi$ & $a_1(C_{{ud,s}}-C_{{us,d}})$ \\
$\Sigma^0\to \psi \gamma$ & $\frac{b_1 (C_{us,d}-C_{ud,s})}{\sqrt{2}}$ & $\Xi^0\to \overline K^0 \psi$ & $(a_1+a_2)(C_{{ud,s}}-C_{{us,d}})$ \\
$\Xi^0\to \psi\gamma$ & $b_1 C_{us,s}$ & $n\to \pi^0 \psi$ & $\frac{a_2C_{{ud,d}}}{\sqrt{2}}$ \\
$n\to \psi\gamma$ & $-b_1 C_{ud,d}$ & $n\to K^0 \psi$ & $(a_1+a_2)C_{{ud,s}}$ \\ \hline
\end{tabular}
\end{table}
\section{Effective Lagrangian analysis}
\label{sec:EFT}
The effective Lagrangian in Eq.~\ref{eq:Lagrangian_eff} can be matched onto the corresponding chiral effective theory.  At the leading order,
the chiral representation of the effective Lagrangian  can be expressed as~\cite{Alonso-Alvarez:2021oaj},
\begin{eqnarray}
	\mathcal{L}^{(0)}_{\rm{cheff}} =
	\alpha\Big\langle\hat{C}^{L}u^\dagger\mathcal{B}_{\mathcal{R}}\psi_{\mathcal{R}}u^\dagger\Big\rangle
	+
	\beta\Big\langle\hat{C}^{R}u^\dagger\mathcal{B}_{\mathcal{R}}\psi_{\mathcal{R}}u \Big\rangle\,,
	\label{eq:LeffChPT}
\end{eqnarray}
here \(\alpha\) and \(\beta\) are nonperturbative low-energy constants, and $u$ is described by the pseudo-Goldstone boson field $\mathcal{P}$, with $u^2=\exp\!\left(\frac{i\,\mathcal{P}}{f_{\pi}}\right)$. The coefficients $C^{L/R}_{ij,k}$ of the effective Lagrangian are treated as spurion fields $\hat{C}^{L/R}$, defining $\hat{C}^{L/R}\equiv \frac{1}{2}\sum_{i,j} \epsilon_{mij}C_{ij,n}^{L/R}$. The Lagrangian can be expanded into a meson-free term $\mathcal{L}^0$ that couples only light baryons and dark baryons, and a one-meson term $\mathcal{L}^1$ that couples the light baryon and dark baryon to a single meson.
\begin{eqnarray}
\mathcal{L}^0&=&\alpha\left\langle\hat C^{L}\mathcal{B}_{\mathcal{R}}\right\rangle\psi_{\mathcal{R}}+\beta\left\langle\hat C^{R}\mathcal{B}_R\right\rangle\psi_{\mathcal{R}},\\
\mathcal{L}^1&=&-\frac{i\alpha}{2f_{\pi}}\left\langle\hat C^{L}\{\mathcal{P},\mathcal{B}_R\}\right\rangle \psi_{\mathcal{R}}-\frac{i\beta}{2f_{\pi}}\left\langle\hat C^{R}[\mathcal{P},\mathcal{B}_R]\right\rangle\psi_{\mathcal{R}}.
\end{eqnarray}
here, we take $\alpha=-\beta=-0.014(2)$ GeV$^3$, $f_{\pi}=0.131$ GeV. The primary focus of our work is the vector mediator model $X_{\mu}$. Flavor mixing effects in the neutral meson system arising from the exchange of the color-triplet vector lead to stringent bounds on the couplings $C^{L/R}$~\cite{Alonso-Alvarez:2021oaj,Alonso-Alvarez:2021qfd},
\begin{align}
	C^L_{ud,d}<0.09\,{\rm TeV}^{-2},\
	C^L_{ud,s}<0.09\,{\rm TeV}^{-2},\
	C^L_{us,d}<0.08\,{\rm TeV}^{-2},\
	C^L_{us,s}<0.08\,{\rm TeV}^{-2}.
\end{align}
We will use an experimental upper constraint on $C^L_{us,s}<5.5\times 10^{-2}$ TeV$^{-2}$($m_{\psi}=1.07$ GeV) which coming from recent BESIII experiment on the semi-invisible decays of $\Xi^-\to \pi^- \psi$~\cite{BESIII:2025sfl}.
In the following,  we will focus on the possible semi-invisible decays of hyperons, including invisible decays into pion plus dark baryon (hadronic channel), and into photon plus dark baryon (radiative channel). In addition, we will also discuss the analogous radiative decays of neutron into a photon and an invisible dark baryon.
\subsection{Hyperon decays into $\pi$ and $\psi$}
\begin{figure}[H]
	\centering
	{\includegraphics[width=1\textwidth]{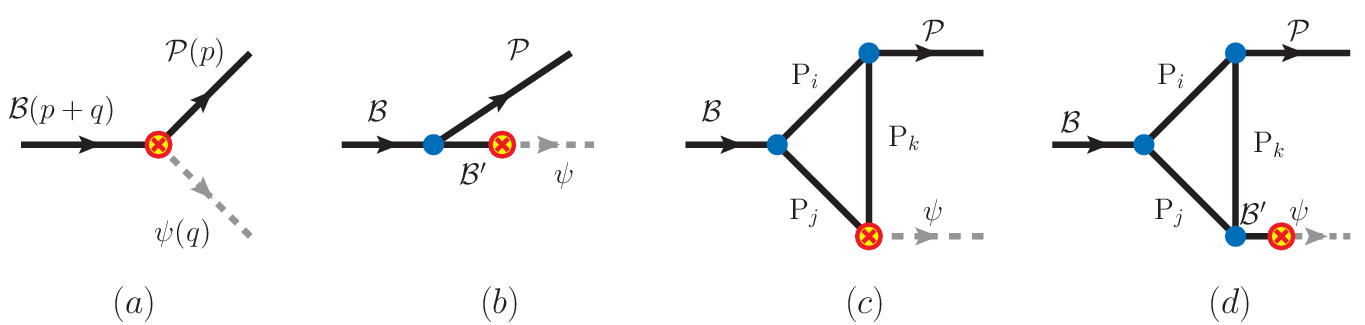}}
	\caption{The tree and triangle diagrams of semi-invisible hyperon decays in the effective Lagrangian framework. Diagrams (a) and (b) provide the leading-order contributions, while (c) and (d) are next-to-leading-order ones.}
	\label{fig:loop}
\end{figure}

The possible hadronic diagrams for hyperon decays into $\pi$ and $\psi$ are shown in Fig.~\ref{fig:loop}. The vertices of the first two diagrams are primarily determined by $\mathcal{L}^0$  and $\mathcal{L}^1$, whereas the two triangle diagrams account for possible hadron exchange processes at the one-loop level. For Fig.~\ref{fig:loop}(c), the intermediate exchanged particles (${\rm P_i},{\rm P_j};{\rm P_k}$) can be either ($\mathcal{V},\mathcal{B};\mathcal{P}$) or ($\mathcal{B},\mathcal{P};\mathcal{B}$), for Fig.~\ref{fig:loop}(d), they (${\rm P_i},{\rm P_j};{\rm P_k}, \mathcal{B}'$) are  ($\mathcal{B},\mathcal{P}/\mathcal{V},\mathcal{B}$) or ($\mathcal{P},\mathcal{B},\mathcal{V}$). The relevant effective Lagrangian is as follows~\cite{deSwart:1963pdg,Ronchen:2012eg,Oh:2004wp,Xing:2023kjk,Hu:2024uia}.
\begin{eqnarray}
	\mathcal{L}_{\mathcal{B}\mathcal{B}\mathcal{P}} &=&  g_{\mathcal{B}\mathcal{B}\mathcal{P}} \bar{\mathcal{B}} \gamma_5 \mathcal{P} \mathcal{B},	\label{eq:L_BBP_trace}\\
	\mathcal{L}_{\mathcal{B'}\mathcal{B}\mathcal{V}} &=&-i g_{\mathcal{B'}\mathcal{B}\mathcal{V}} \bar{\mathcal{B}'}\gamma_\mu \mathcal{V}^\mu \mathcal{B} - \frac{f_{\mathcal{B'}\mathcal{B}\mathcal{V}}}{m_{\mathcal{B'}}+m_\mathcal{B}'} \bar{\mathcal{B}'}\sigma_{\mu\nu}\partial^\nu \mathcal{V}^\mu \mathcal{B} \label{eq:LBBV}\,,\\
	\mathcal{L}_{\mathcal{V}\mathcal{P}\mathcal{P}} &=& g_{\mathcal{V}\mathcal{P}\mathcal{P}} \mathcal{V}^\mu[\mathcal{P},\partial_\mu \mathcal{P}] \label{eq:VPP}\,,
\end{eqnarray}
where $\mathcal{P}$, $\mathcal{V}^\mu$, and $\mathcal{B}$ denotes the pseudoscalar-octet meson, vector meson and light baryon octet, respectively. The couplings $g_{\mathcal{BBV}}$,$f_{\mathcal{BBV}}$ and $g_{\mathcal{VPP}}$~\cite{Aliev:2009ei,Aliev:2006xr} are collected in Tab.\ref{tab:coupling}.


The dominant amplitude for the semi-invisible decay of hyperons is as follows.
\begin{eqnarray}
	&\mathcal{M}_{\Sigma^+\to\pi^+\psi}&=\mathcal{T}_a(\Sigma^+,\pi^+)+\mathcal{T}_b(\Sigma^+,\Lambda/\Sigma^0,\pi^+)+
	\mathcal{M}_c(\rho,\Lambda/\Sigma;\pi)+\mathcal{M}_c(K^{*},\Xi/p;K)\notag\\
	&&+\mathcal{M}_c(\Lambda/\Sigma,\pi/\eta_8;\Lambda/\Sigma)+\mathcal{M}_c(\Xi^0,K^+;\Xi^-)+\mathcal{M}_c(p,\overline K^0;n)+\mathcal{M}_d(\rho,\Lambda/\Sigma;\pi,\Lambda/\Sigma)
	\notag\\
	&&+\mathcal{M}_d(K^{*},\Xi/p;K,\Lambda/\Sigma)+\mathcal{M}_d(\Lambda/\Sigma,\pi/\eta_8;\Lambda/\Sigma,\Lambda/\Sigma)+\mathcal{M}_d(\Xi^0,K^+;\Xi^-,\Lambda/\Sigma^0)\notag\\
	&&+\mathcal{M}_d(p,\overline K^0;n,\Lambda/\Sigma^0),
	\notag\\
	&\mathcal{M}_{\Sigma^-\to\pi^-\psi}&=\mathcal{T}_a(\Sigma^-,\pi^-)+\mathcal{T}_b(\Sigma^-,\Lambda/\Sigma^0,\pi^-)+
	\mathcal{M}_c(\rho,\Lambda/\Sigma;\pi)+\mathcal{M}_c(K^{*},\Xi/n;K)\notag\\
	&&+\mathcal{M}_c(\Lambda/\Sigma,\pi/\eta_8;\Lambda/\Sigma)+\mathcal{M}_c(\Xi^-,K^0;\Xi^0)+\mathcal{M}_c(n,K^-;p)+\mathcal{M}_d(\rho,\Lambda/\Sigma;\pi,\Lambda/\Sigma^0)
	\notag\\
	&&+\mathcal{M}_d(K^{*},\Xi^-/n;K,\Lambda/\Sigma^0)+\mathcal{M}_d(\Lambda/\Sigma,\pi/\eta_8;\Lambda/\Sigma,\Lambda/\Sigma^0)+\mathcal{M}_d(\Xi^-,K^0;\Xi^0,\Lambda/\Sigma^0)
	\notag\\
	&&+\mathcal{M}_d(n,K^-;p,\Lambda/\Sigma^0),
	\nonumber\\
	&\mathcal{M}_{\Xi^-\to\pi^-\psi}&=\mathcal{T}_a(\Xi^-,\pi^-)+\mathcal{T}_b(\Xi^-,\Xi^0,\pi^-)+\mathcal{M}_c(K^{*-},\Lambda/\Sigma^0;\overline K^0)+\mathcal{M}_c(\rho,\Xi;\pi)
	\notag\\
	&&
	+\mathcal{M}_c(\Lambda/\Sigma,K;\Lambda/\Sigma)+\mathcal{M}_c(\Xi^-,\pi^0/\eta_8;\Xi^0)+\mathcal{M}_d(\rho,\Xi;\pi,\Xi^0)+\mathcal{M}_d(K^{*},\Lambda/\Sigma^0; K,\Xi^0)
	\notag\\
	&&+\mathcal{M}_d(\Lambda/\Sigma,K;\Lambda/\Sigma,\Xi^0)+\mathcal{M}_d(\Xi^-,\pi^0/\eta_8;\Xi^0,\Xi^0),
	\nonumber\\
	&\mathcal{M}_{\Lambda\to\pi^0\psi}&=\mathcal{T}_a(\Lambda,\pi^0)+\mathcal{T}_b(\Lambda,\Lambda/\Sigma^0,\pi^0)+
	\mathcal{M}_c(\rho,\Sigma,\pi)+\mathcal{M}_c(K^{*},p/n/\Xi,K)
	\notag\\
	&&+\mathcal{M}_c(\Lambda/\Sigma,\pi/\eta_8;\Lambda/\Sigma)+\mathcal{M}_c(p/n,K;p/n)+\mathcal{M}_c(\Xi,K;\Xi)+\mathcal{M}_d(\rho,\Sigma;\pi,\Lambda/\Sigma^0)
	\notag\\
	&&+\mathcal{M}_d(K^{*},p/n/\Xi;K,\Lambda/\Sigma^0)+\mathcal{M}_d(\Lambda/\Sigma,\pi/\eta_8;\Lambda/\Sigma,\Lambda/\Sigma^0)+\mathcal{M}_d(p/n,K;p/n,\Lambda/\Sigma^0)
	\notag\\
	&&+\mathcal{M}_d(\Xi,K;\Xi,\Lambda/\Sigma^0),
	\notag\\
	%
	%
	&\mathcal{M}_{\Xi^0\to\pi^0\psi}&=\mathcal{T}_a(\Xi^0,\pi^0)+\mathcal{T}_b(\Xi^0,\Xi^0,\pi^0)+
	\mathcal{M}_c(K^{*},\Lambda/\Sigma;K)+\mathcal{M}_c(\rho^+,\Xi^-;\pi^+)
	\notag\\
	&&+\mathcal{M}_c(\Lambda/\Sigma,K;\Lambda/\Sigma)+\mathcal{M}_c(\Xi,\pi/\eta_8;\Xi)+\mathcal{M}_d(\rho^+,\Xi^-;\pi^+,\Xi^0)
	+\mathcal{M}_d(K^{*},\Lambda/\Sigma;K,\Xi^0)
	\notag\\
	&&+\mathcal{M}_d(\Lambda/\Sigma,K;\Lambda/\Sigma,\Xi^0)
	+\mathcal{M}_d(\Xi,\pi/\eta_8;\Xi,\Xi^0).
\end{eqnarray}
Here, $\mathcal{T}_a(\mathcal{B},\pi)$ and $\mathcal{T}_b(\mathcal{B},\mathcal{B}^{\prime},\pi)$ are tree progresses, while $\mathcal{M}_c({\rm P_i},{\rm P_j};{\rm P_k})$ and $\mathcal{M}_d({\rm P_i},{\rm P_j};{\rm P_k},\mathcal{B})$ represent one-loop progresses. Their explicit forms are given in the Appendix.
\begin{table}
	\centering
	\renewcommand{\arraystretch}{1}
	\setlength{\tabcolsep}{10pt}
	\caption{The $\mathcal{BBP}$~\cite{Aliev:2006xr} , $\mathcal{BBV}$~\cite{Aliev:2009ei} and $\mathcal{VPP}$~\cite{Ronchen:2012eg} strong coupling constants used in this work.}
	\label{tab:coupling}
	\begin{tabular}{ccccccc}
		\hline\hline
		$g_{\Sigma\Xi K}$ & $g_{\Lambda\Sigma\pi}$ & $g_{\Lambda\Xi K}$ & $g_{NN\pi}$ & $g_{N\Sigma K}$ & $g_{\Sigma N K}$ & $g_{\Sigma\Sigma\pi}$ \\
		$13\pm3$ & $10\pm3$ & $4.5\pm2$ & $14\pm4$ & $3.2\pm2.2$ & $4\pm3$ & $9\pm2$ \\
		
		$g_{\Xi\Sigma K}$ & $g_{\Xi\Xi\pi}$ & $g_{\Sigma\Lambda\rho}$ & $g_{\Sigma\Sigma\rho}$ & $g_{\Sigma\Sigma\phi}$ & $g_{\Xi\Xi\rho}$ & $g_{\Xi\Xi\phi}$ \\
		$12.5\pm3$ & $10\pm2$ & $2\pm0.6$ & $7.2\pm1.2$ & $-6\pm0.8$ & $4.2\pm2.1$ & $-9.5\pm2.5$ \\
		
		$g_{\Lambda\Lambda\omega}$ & $g_{\Lambda\Lambda\phi}$ & $g_{\Xi\Sigma K^*}$ & $g_{NN\omega}$ & $g_{N\Lambda K^*}$ & $g_{N\Sigma K^*}$ & $g_{\Xi\Lambda K^*}$ \\
		$-7.1\pm1.1$ & $-5.3\pm1.5$ & $2.3\pm1.7$ & $-8.9\pm1.5$ & $5.1\pm1.8$ & $4\pm0.7$ & $-5.9\pm0.7$ \\
		
		$g_{NN\rho}$ & $f_{\Sigma\Lambda\rho}$ & $f_{\Sigma\Sigma\rho}$ & $f_{\Sigma\Sigma\phi}$ & $f_{\Xi\Xi\rho}$ & $f_{\Xi\Xi\phi}$ & $f_{\Lambda\Lambda\omega}$ \\
		$2.5\pm1.1$ & $12.3\pm2.3$ & $25\pm1$ & $2.5\pm1.7$ & $1.4\pm0.5$ & $32.3\pm3.9$ & $8.7\pm0.5$ \\
		
		$f_{\Lambda\Lambda\phi}$ & $f_{NN\rho}$ & $f_{NN\omega}$ & $f_{N\Lambda K^*}$ & $f_{N\Sigma K^*}$ & $f_{\Xi\Lambda K^*}$ & $f_{\Xi\Sigma K^*}$ \\
		$24.6\pm3.5$ & $22.2\pm1.7$ & $23.4\pm1.1$ & $-28\pm2.4$ & $-1.2\pm1.1$ & $17.5\pm2.2$ & $36.1\pm3.2$ \\
		
		$g_{\rho\pi\pi}$ & $g_{K^*K\pi}$ &  &  &  &  &  \\
		$4.2\pm0.4$ & $3\pm0.3$ &  &  &  &  &  \\
		\hline
	\end{tabular}
\end{table}
\subsection{Hyperon decays into $\psi$ and $\gamma$}
\begin{figure}[H]
	\centering
	{\includegraphics[width=0.8\textwidth]{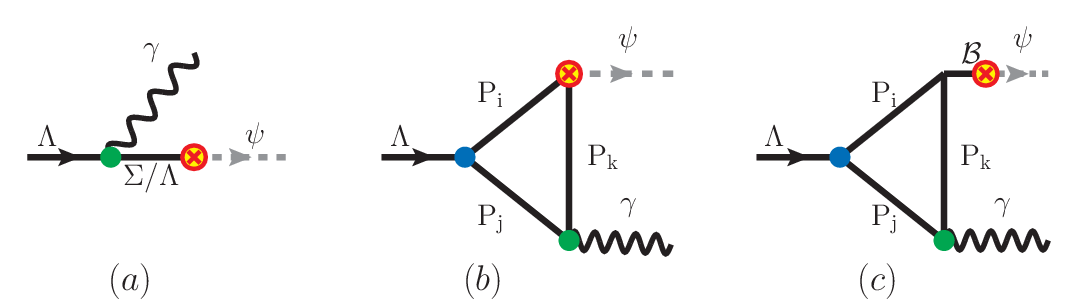}}
	\caption{The leading order radiative semi-invisible decay of hyperons is shown in diagram (a), while (b) and (c) are the next-to-leading order triangle diagram contributions.}
	\label{fig:rad}
\end{figure}

We present the Feynman diagrams for the radiative semi-invisible decay of hyperons and neutron in Fig.~\ref{fig:rad}. The intermediate particles (${\rm P_i},{\rm P_j},{\rm P_k}$) can be ($\mathcal{P}/\mathcal{V},\mathcal{B},\mathcal{B}$) in Fig.~\ref{fig:rad}(b), and ($\mathcal{B},\mathcal{P}/\mathcal{V},\mathcal{P}/\mathcal{V}$) in Fig.~\ref{fig:rad}(c). Clearly, at leading order, this process requires the radiative decays of light hadrons. The SU(3) invariant Lagrangians for the hadron-photon coupling are~\cite{Suo:2025rty,Ren:2022fhp,Gao:2010hy}
\begin{eqnarray}
&&\mathcal{L}_{\gamma \mathcal{P}_1\mathcal{P}_2}=i e A_{\mu} (\mathcal{P}_1\partial^{\mu} \mathcal{P}_2-\partial^{\mu} \mathcal{P}_1 \mathcal{P}_2),\\
&&\mathcal{L}_{\gamma \mathcal{P}\mathcal{V}}=\frac{e}{m_{\mathcal{P}}} g_{\gamma \mathcal{P}\mathcal{V}} \varepsilon^{\mu\nu\alpha\beta}\partial_{\mu} \mathcal{V}_{\nu} \partial_{\alpha} A_{\beta} \mathcal{P},\\
&&\mathcal{L}_{\gamma \mathcal{B}_8\mathcal{B}_8}=- e \overline {\mathcal{B}_8} (\gamma^{\mu} A_{\mu} Q_{\mathcal{B}_8}-\frac{\kappa_{\mathcal{B}_8}}{2M_{\mathcal{B}_8}}\sigma^{\mu\nu}\partial_{\nu}A_{\mu})\mathcal{B}_8,\\
&&\mathcal{L}_{\gamma \Lambda \Sigma}=\frac{e \mu_{\Lambda\Sigma} }{2 m_N} \bar \Sigma \sigma_{\mu\nu} \partial^{\nu} A^{\mu} \Lambda+h.c.,
\end{eqnarray}
where $Q_{\mathcal{B}_8}$ is the electric charge, and $\kappa_{\mathcal{B}_8}$ denotes the anomalous magnetic moment of baryon. In the paper, we take $\kappa_n=-1.913$, $\kappa_p=1.793$, $\kappa_{\Sigma^+}=1.458$, $\kappa_{\Sigma^-}=-0.16$~\cite{deSwart:1963pdg}, and transition magnetic moment $\mu_{\Lambda\Sigma}=-1.61$~\cite{Huang:2021ahp}. Here we do not include the contributions from flavor-changing electro-weak decays, such as $\Xi^0\to \Lambda \gamma$, they are largely suppressed and have very small effect in our processes.

The amplitudes for the radiation semi-invisible decay of hyperons are given as,
\begin{eqnarray}
	&\mathcal{M}_{\Lambda\to\gamma\psi}&=\mathcal{T}^{\prime}_a(\Lambda,\Lambda/\Sigma^0)+\mathcal{M}^{\prime}_b(\Sigma,\pi;\pi)+\mathcal{M}^{\prime}_b(\Xi/p,K;K)
	+\mathcal{M}^{\prime}_b(\pi/\eta_8,\Lambda/\Sigma;\Lambda/\Sigma)
	\notag\\
	&&+\mathcal{M}^{\prime}_b(K,p/n/\Xi;p/n/\Xi)+\mathcal{M}^{\prime}_b(\Lambda/\Sigma,\rho/\omega/\phi;\pi/\eta_8)+\mathcal{M}^{\prime}_b(p/n/\Xi,K^{*};K)
	\notag\\
	&&+\mathcal{M}^{\prime}_c(K,p/n/\Xi;p/n/\Xi,\Lambda/\Sigma^0)+\mathcal{M}^{\prime}_c(\Xi^-/p,K;K,\Lambda/\Sigma^0)+\mathcal{M}^{\prime}_c(\pi/\eta,\Lambda/\Sigma;\Lambda/\Sigma,\Lambda/\Sigma)
	\notag\\
	&&+\mathcal{M}^{\prime}_c(\Lambda/\Sigma,\rho/\omega/\phi;\pi/\eta,\Lambda/\Sigma^0)+\mathcal{M}^{\prime}_c(p/n/\Xi,K^{*};K,\Lambda/\Sigma^0)+\mathcal{M}^{\prime}_c(\Sigma,\pi;\pi,\Lambda/\Sigma^0),\notag\\
	&\mathcal{M}_{\Sigma^0\to\gamma\psi}&=\mathcal{T}^{\prime}_a(\Sigma^0,\Sigma^0/\Lambda)+\mathcal{M}^{\prime}_b(\Sigma,\pi;\pi)+\mathcal{M}^{\prime}_b(p/\Xi,K;K)
	+\mathcal{M}^{\prime}_b(\pi/\eta_8,\Lambda/\Sigma;\Lambda/\Sigma)
	\notag\\
	&&+\mathcal{M}^{\prime}_b(K,p/n/\Xi;p/n/\Xi)+\mathcal{M}^{\prime}_b(\Lambda/\Sigma,\rho/\omega/\phi;\pi/\eta_8)+\mathcal{M}^{\prime}_b(p/n/\Xi,K^{*};K)\notag\\
	&&+\mathcal{M}^{\prime}_c(\Xi^-/p,K;K,\Lambda/\Sigma^0)+\mathcal{M}^{\prime}_c(\pi/\eta,\Lambda/\Sigma;\Lambda/\Sigma,\Lambda/\Sigma^0)+\mathcal{M}^{\prime}_c(K,p/n/\Xi;p/n/\Xi,\Lambda/\Sigma^0)
	\notag\\
	&&+\mathcal{M}^{\prime}_c(\Lambda/\Sigma,\rho/\omega/\phi;\pi/\eta,\Lambda/\Sigma^0)+\mathcal{M}^{\prime}_c(p/n/\Xi,K^{*};K,\Lambda/\Sigma^0)+\mathcal{M}^{\prime}_c(\Sigma,\pi;\pi,\Lambda/\Sigma),\notag\\
	&\mathcal{M}_{\Xi^0\to\gamma\psi}&=\mathcal{T}^{\prime}_a(\Xi^0,\Xi^0)+\mathcal{M}^{\prime}_b(\Sigma^+,K^-;K^+)+\mathcal{M}^{\prime}_b(\Xi^-,\pi^+;\pi^-)
	+\mathcal{M}^{\prime}_b(K,\Lambda/\Sigma;\Lambda/\Sigma)\notag\\
	&&+\mathcal{M}^{\prime}_b(\pi/\eta_8,\Xi;\Xi)+\mathcal{M}^{\prime}_b(\Lambda/\Sigma,K^{*};K)+\mathcal{M}^{\prime}_b(\Xi,\rho/\omega/\phi;\pi/\eta_8)+\mathcal{M}^{\prime}_c(\pi/\eta,\Xi;\Xi,\Xi^0)\notag\\
	&&+\mathcal{M}^{\prime}_c(K,\Lambda/\Sigma;\Lambda/\Sigma,\Xi^0)+\mathcal{M}^{\prime}_c(\Lambda/\Sigma,K^{*};K,\Xi^0)+\mathcal{M}^{\prime}_c(\Xi,\rho/\omega/\phi;\pi/\eta,\Xi^0)\notag\\
	&&+\mathcal{M}^{\prime}_c(\Xi^-,\pi^+;\pi^-,\Xi^0)+\mathcal{M}^{\prime}_c(\Sigma^+,K^-;K^+,\Xi^0).
\end{eqnarray}
Assuming that the neutron can decay into a dark baryon and that the effective operators are identical, we also present the possible decay amplitude,
\begin{eqnarray}
&\mathcal{M}_{n \to\gamma\psi}&=\mathcal{T}^{\prime}_a(n,n)+\mathcal{M}^{\prime}_b(p,\pi^-;\pi^+)+\mathcal{M}^{\prime}_b(\Sigma^-,K^+;K^-)+\mathcal{M}^{\prime}_b(\pi/\eta_8,p/n;p/n)
\notag\\
&&+\mathcal{M}^{\prime}_b(K,\Lambda/\Sigma;\Lambda/\Sigma)+\mathcal{M}^{\prime}_b(p/n,\rho/\omega;\pi/\eta_8)+\mathcal{M}^{\prime}_b(\Lambda/\Sigma,K^{*};K)\notag\\
&&+\mathcal{M}^{\prime}_c(p,\pi^-;\pi^+,n)+\mathcal{M}^{\prime}_c(\Sigma^-,K^+;K^-,n)+\mathcal{M}^{\prime}_c(\pi/\eta_8,p/n;p/n,n)\notag\\
&&+\mathcal{M}^{\prime}_c(K,\Lambda/\Sigma;\Lambda/\Sigma,n)+\mathcal{M}^{\prime}_c(p/n,\rho/\omega;\pi/\eta_8,n)+\mathcal{M}^{\prime}_c(\Lambda/\Sigma,K^{*};K,n).
\end{eqnarray}
Note that owing to phase-space and experimental constraints, when neutrons can undergo semi-invisible decays, the relevant Wilson coefficient $C^L_{ud,d}$ should be constrained to a lower value.
\section{Numerical analysis: branching ratios}
\label{sec:analysis}
For the branching ratio of invisible hyperon decays, within the kinematically allowed region, it depends strongly on the mass of the dark baryon $m_{\psi}$. Although the dark baryon decay mechanisms based on neutrons or b-hadrons may impose strong constraints on the mass, for a fuller consideration here, we adopt the dark baryon mass within the kinematic range of the hyperon:
\begin{eqnarray}\label{eq:mpsi}
0.94\ \rm{GeV}< m_{\psi} <1.18\ \rm{GeV},
\end{eqnarray}
In this region, all hyperons may decay into a dark baryon $\psi$ and $\pi$ or $\gamma$. If kinematically allowed, the semi-invisible decay of the neutron is also possible, and we will then briefly discuss it.

Triangle loop diagrams involve final-state strong interactions among hadrons. To account for the internal structure of hadrons as well as off-shell effects, phenomenological form factors should be introduced as~\cite{Li:1996yn,Buzatu:1993xu},
\begin{equation}
\mathcal F(k^2,m_{E}^2) = \Big(\frac{m_E^2-\Lambda^2}{k^2-\Lambda^2}\Big)^2,
\label{eq:monopole_ff}
\end{equation}
where $k$ and $m_E$ are the momentum and mass of the exchanged particle. $\Lambda$ denotes the cutoff parameter parameterized as $\Lambda=m_E+\eta \Lambda_{QCD}$, here parameter taken $1\leq\eta\leq2$, and the QCD scale $\Lambda_{QCD}=0.33$ GeV~\cite{Cheng:2004ru,Han:2021azw}.

For the semi-invisible decays of hyperons, after including the contributions from tree-level and triangular hadronic loop diagrams, the resulting branching ratios are presented in Tab.~\ref{tab:brtab}. A dark baryon mass of $1.0$ GeV is adopted, and the uncertainties coming from relevant effective coupling coefficients $C^{L}$ and strong coupling constants. For comparison, we also present in the table the experimental constraints~\cite{Bollig:2020xdr,BESIII:2021slv}.

\begin{figure}[H]
	\centering
	{\includegraphics[scale=1,width=1\textwidth]{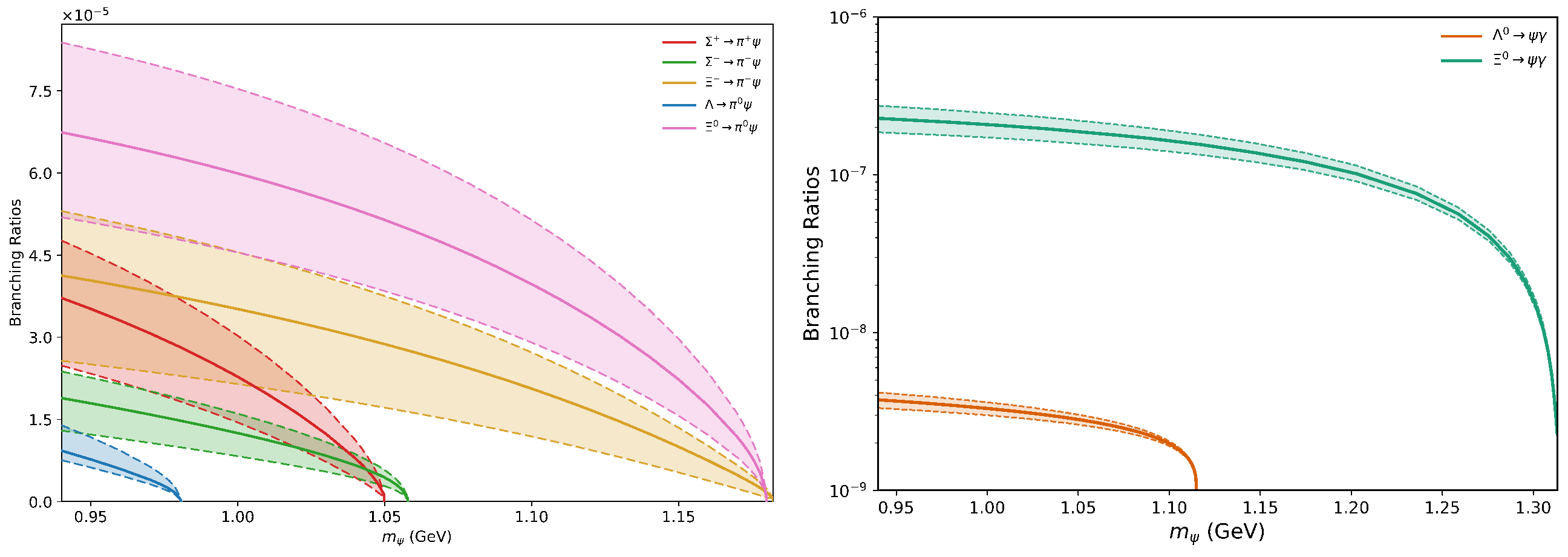}}
	\caption{The branching ratios of the hadronic semi-invisible modes (left)
		$\Sigma^+\to\pi^+\psi$, $\Sigma^-\to\pi^-\psi$, $\Xi^-\to\pi^-\psi$,
		$\Lambda^+\to \pi^0\psi$, $\Sigma^0\to \pi^0\psi$, $\Xi^0\to \pi^0\psi$, and the radiative semi-invisible modes (right) $\Lambda\to \gamma \psi$, $\Xi^0\to \gamma\psi$
		as functions of the dark-fermion mass $m_\psi$. The solid curves denote the central values, while the shaded bands bounded by dashed curves show the estimated uncertainties.}
	\label{fig:BRmpsi}
\end{figure}

\begin{figure}[H]
	\centering
	{\includegraphics[width=0.98\textwidth]{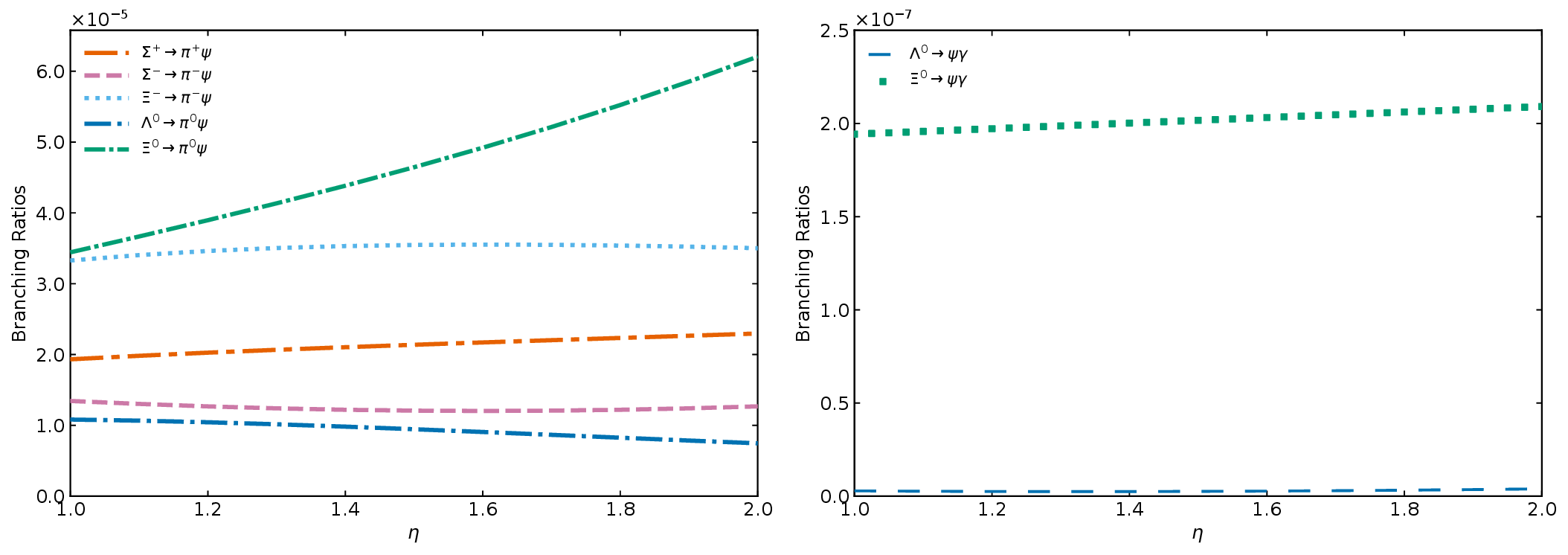}}
	\caption{Dependence of branching ratios on $\eta$ at $m_{\psi}=1.00$ GeV, including the hadronic (left) and radiative (right) semi-invisible hyperon decay modes.}
	\label{fig:BReta}
\end{figure}

Our calculations indicate that the Wilson coefficients and tree diagram contributions dominate the final branching ratios. The triangle diagrams, originating from rescattering effects of final-state particles, involve strong-interaction vertices despite being loop processes, thereby providing substantial contributions. For the Wilson coefficients, we incorporate constraints from flavor mixing in neutral meson systems along with the upper limits from BESIII measurements. Consequently, for $m_{\psi}=1.0$ GeV, the semi-invisible decay branching ratios of hyperons are found to be of order $10^{-5}$.
Even though the tree-level contributions are dominant, one-loop triangle contribution still yield significant corrections for a dark matter mass near 1.0 GeV. For instance, in processes $\Sigma^-\to\pi^-\psi$, $\Xi^0\to\pi^0\psi$ and $\Lambda\to\pi^0\psi$, the branching ratios of the one-loop triangle diagrams in Fig.~\ref{fig:loop}.(c,d) alone are comparable to those of the tree diagrams in Fig.~\ref{fig:loop}.(a,b).
\begin{eqnarray}
&\Sigma^-\to\pi^-\psi:&\ \mathcal{B}r(a,b)=1.83\times10^{-5},\ \mathcal{B}r(c,d)=0.94\times10^{-5},\ \mathcal{B}r(a,b,c,d)=1.25\times10^{-5},\ \notag\\
&\Xi^0\to\pi^0\psi:&\ \mathcal{B}r(a,b)=5.24\times10^{-5},\ \mathcal{B}r(c,d)=1.05\times10^{-5},\ \mathcal{B}r(a,b,c,d)=5.99\times10^{-5},\ \notag\\
&\Lambda\to\pi^0\psi:&\ \mathcal{B}r(a,b)=1.93\times10^{-5},\ \mathcal{B}r(c,d)=0.49\times10^{-5},\ \mathcal{B}r(a,b,c,d)=0.77\times10^{-5}.
\end{eqnarray}
Given the non-negligible contribution from loop diagrams, we then examine the dependence of the branching fraction on the dark-baryon mass $m_{\psi}$ and cutoff parameter $\eta$. The figures are respectively shown in Fig.~\ref{fig:BRmpsi} and Fig.~\ref{fig:BReta}, in which the branching ratios decrease with increasing $m_{\psi}$. At the minimum allowed value of $m_{\psi}=0.94$ GeV in Eq.~\ref{eq:mpsi}, the process $\Xi^0\to\pi^0\psi$ has the largest branching ratio, $\mathcal{B}r(\Xi^0\to\pi^0\psi)=7.0\times10^{-5}$, whereas $\Lambda\to \pi^0\psi$ yields the smallest $\mathcal{B}r(\Lambda\to \pi^0\psi)=1.16\times10^{-5}$. In addition, the branching ratios do not vary significantly with $\eta$, except for processes $\Xi^0 \to \pi^0\psi$ and $\Sigma^-\to \pi^-\psi$, which increase strongly with $\eta$. This may be driven by the large loop contributions in these two processes.

In contrast to previous SU(3) analyses, the calculation from the effective Lagrangian show certain deviations, suggesting that flavor-symmetry breaking effects are sizable in these semi-invisible decays.
\begin{align}
R_{\Xi}^{\text{SU3}} &= \frac{\Gamma(\Xi^-\to \pi^- \psi)}{\Gamma(\Xi^0\to \pi^0\psi)} = 2.00, \quad
R_{\Xi}^{\text{eff}} = \frac{\Gamma(\Xi^-\to \pi^- \psi)}{\Gamma(\Xi^0\to \pi^0\psi)} = 1.04, \notag\\
R_{\Sigma}^{\text{SU3}} &= \frac{\Gamma(\Sigma^-\to\pi^- \psi)}{\Gamma(\Xi^0\to \pi^0\psi)} = 0.07, \quad
R_{\Sigma}^{\text{eff}} = \frac{\Gamma(\Sigma^-\to\pi^- \psi)}{\Gamma(\Xi^0\to \pi^0\psi)} = 0.40, \notag\\
R_{\Lambda}^{\text{SU3}} &= \frac{\Gamma(\Lambda\to \gamma \psi)}{\Gamma(\Xi^0\to \gamma\psi)} = 0.35, \quad
R_{\Lambda}^{\text{eff}} = \frac{\Gamma(\Lambda\to \gamma \psi)}{\Gamma(\Xi^0\to \gamma\psi)} = 0.02.
\end{align}
These difference may stem from phase space and dynamical effects of loop triangle diagrams. For instance, the tree-level ratio $\widetilde{R}_{\Xi}^{eff}=2.03$ is in good agreement with the SU(3) prediction. However, the final-state interaction effect, as well as the interference between tree and loop amplitudes leads to large deviation, when including the triangle one-loop diagram contributions.

If phase space permits, neutrons may also undergo the radiative semi-invisible decay process. This provide a possible explanation for the existing discrepancy in neutron lifetime measurements between the bottle and beam experiments~\cite{Yue:2013qrc,Serebrov:2004zf}. However, the discrepancy between the two experimental methods regarding the branching ratio is only 1\%, which strongly constrains the decay width $\Gamma(n\to \gamma\psi)=7.41\times 10^{-30}$. Accordingly, using the width together with the effective Lagrangian analysis, we obtain an new upper limit of $C_{ud,d}^{L}=6.87\times 10^{-11}$ GeV$^{-2}$.

\begin{table}
	\centering
	\caption{The branching ratios of light baryon decays into light meson and $\psi$ within $m_{\psi}=1.00$ GeV (for $\Lambda\to \pi^0\psi$, we take $m_{\psi}=0.95$ GeV). The uncertainty analysis mainly comes from couplings and coefficients $C^{L/R}_{ud/s,d/s}$. Tree represents the contribution of tree diagram, while FSI represents  contribution from the triangle diagrams.}
	\label{tab:brtab}
	\begin{tabular}{c|c|c|c|c}
		\hline\hline
		\multirow{2}{*}{channel} & \multicolumn{3}{c|}{Branching ratios }  &\multirow{2}{*}{exp}\\\cline{2-4}
                                             & Tree & FSI & Total   &
		\\
		\hline
		$\Sigma^+ \to \pi^+ \psi$  & $2.60\times 10^{-5}$ & $0.07\times 10^{-5}$ & $2.28^{+0.75}_{-0.83}\times 10^{-5}$& $<10^{-4}$~\cite{Bollig:2020xdr}\\
		$\Sigma^- \to \pi^- \psi$  & $1.83\times 10^{-5}$ & $0.94\times 10^{-5}$ & $1.25^{+0.35}_{-0.42}\times 10^{-5}$ & $< 10^{-4}$~\cite{Bollig:2020xdr} \\
		$\Sigma^0 \to \pi^0 \psi$  &$1.35\times10^{-14}$& $2.01\times10^{-15}$& $6.08^{+0.42}_{-0.48}\times10^{-15}$  & $-$\\
        $\quad \to \gamma \psi$  & $9.25\times 10^{-18}$ &$4.17\times 10^{-18}$& $2.88^{+0.67}_{-0.48}\times 10^{-17}$  &$-$\\
		$\Xi^- \to \pi^- \psi$  & $6.02\times 10^{-5}$ & $0.73\times 10^{-5}$& $3.51^{+1.04}_{-1.37}\times 10^{-5}$ & $< 10^{-5}$~\cite{BESIII:2025sfl}\\
		$\Xi^0 \to \pi^0 \psi$  &  $5.24\times 10^{-5}$ & $1.05\times 10^{-5}$& $5.99^{+1.54}_{-1.44}\times 10^{-5}$ & $-$\\
		$\quad \to \gamma \psi$ & $0.20\times 10^{-6}$ &  $1.96\times 10^{-9}$ & $0.20^{+0.02}_{-0.03}\times 10^{-6}$ &$-$\\
		$\Lambda^0 \to \pi^0 \psi$  & $0.27\times 10^{-5}$ & $0.06\times 10^{-5}$& $0.07^{+0.03}_{-0.01}\times 10^{-5}$ & $< 10^{-7}$~\cite{Bollig:2020xdr}\\
		$\quad \to \gamma \psi$  & $6.39\times 10^{-9}$ &$8.55\times 10^{-9}$& $3.95^{+0.79}_{-0.03}\times 10^{-9}$  &$-$ \\
        \hline
	\end{tabular}
\end{table}
\section{Conclusion}
\label{sec:conclusion}
In this work, we focus on the semi-invisible decays of hyperons $\mathcal{B}\to \pi\psi$,  within the effective Lagrangian and final-state interaction approach. We calculate the branching ratios of hadronic and electromagnetic semi-invisible decays. The results show that the contribution from triangle-loop diagrams can be the same order with tree-level diagram. The branching ratios of  strange-mesongenesis from hyperons can be the order of $10^{-5}$.

\appendix
\section{Decay amplitudes}
The detailed amplitudes for the semi-invisible decay processes such as $\Sigma^+\to \pi^+\psi$ and $\Lambda\to \pi^0 \gamma$ shown in Fig.~\ref{fig:loop} and Fig.~\ref{fig:rad} are as follows.
\begin{eqnarray}
&\mathcal{T}_a(\Sigma^+,\pi^+)&=\bar{u}_\psi(q)(-i\frac{\alpha\hat C}{2f}P_{\mathcal R})u_{\Sigma}(p+q),\notag\\
&\mathcal{T}_b(\Sigma^+,\Lambda,\pi^+)&= \bar{u}_\psi(q)i(\hat CP_{\mathcal R})\frac{i(\slashed q+m_{\Lambda})}{q^2-m_{\Lambda}^2}(ig_{\Sigma\Lambda\pi}\gamma^5)u_{\Sigma}(p+q),\notag\\
&\mathcal{M}_c(\rho^+,\Lambda;\pi^0)
&=\int\frac{d^4l}{(2\pi)^4}\mathcal{F}(l^2,m^2_{\pi})\bar u_\psi(q)i\left(-i\frac{\alpha\hat C}{2f}P_{\mathcal R}\right)\frac{i(\slashed q-\slashed l+m_{\Lambda})}{(q-l)^2-m_{\Lambda}^2 }
\nonumber\\
&&\frac{i\left[-g_{\mu\nu}+\dfrac{(l+p)_\mu(l+p)_\nu}{m_{\rho^+}^2}\right]}{(l+p)^2-m_{\rho^+}^2 }\left[f_{1,\Sigma^+\Lambda\rho^+}\gamma^\mu-i\frac{f_{2,\Sigma^+\Lambda\rho^+}}{m_{\Sigma^+}+m_{\Lambda}}\sigma^{\mu\rho}(l+p)_\rho\right]
\nonumber\\
&&i\left[g_{\rho^+\pi^+\pi^0}(p-l)^\nu\right]\frac{i}{l^2-m_{\pi^0}^2 }u_{\Sigma^+}(p+q),\nonumber\\
&\mathcal{M}_c(\Sigma^+,\pi^0;\Lambda)
&=\int\frac{d^4l}{(2\pi)^4}\mathcal{F}(l^2,m^2_\Lambda)\bar u_\psi(q)i\left(-i\frac{\alpha\hat C}{2f}P_{\mathcal R}\right)\frac{i(\slashed l+m_{\Lambda})}{l^2-m_{\Lambda}^2 }i(g_{\Sigma^+\Lambda\pi^+}\gamma_5)
\nonumber\\
&&\frac{i(\slashed l+\slashed p+m_{\Sigma^+})}{(l+p)^2-m_{\Sigma^+}^2 }i(g_{\Sigma^+\Sigma^+\pi^0}\gamma_5)u_{\Sigma^+}(p+q)\frac{i}{(q-l)^2-m_{\pi^0}^2 },
\nonumber\\
&\mathcal{M}_d(\rho^+,\Lambda;\pi^0,\Sigma^0)
&=\int\frac{d^4l}{(2\pi)^4}\mathcal{F}(l^2,m_{\pi}^2)\mathcal{F}(l^2,m_{\Sigma}^2)\bar u_\psi(q)i(\hat CP_{\mathcal R})\frac{i(\slashed q+m_{\Sigma^0})}{q^2-m_{\Sigma^0}^2 }i( g_{\Lambda\Sigma^0\pi^0}\gamma_5)
\nonumber\\
&&\frac{i(\slashed q-\slashed l+m_{\Lambda})}{(q-l)^2-m_{\Lambda}^2 }\left[f_{1,\Sigma^+\Lambda\rho^+}\gamma^\mu-i\frac{f_{2,\Sigma^+\Lambda\rho^+}}{m_{\Sigma^+}+m_\Lambda}\sigma^{\mu\rho}(l+p)_\rho\right]
\nonumber\\
&&\frac{i\left[-g_{\mu\nu}+\dfrac{(l+p)_\mu(l+p)_\nu}{m_{\rho^+}^2}\right]}{(l+p)^2-m_{\rho^+}^2 }i\left[g_{\rho^+\pi^+\pi^0}(p-l)^\nu\right]\frac{i}{l^2-m_{\pi^0}^2 }u_{\Sigma^+}(p+q),
\nonumber\\
&\mathcal{M}_d(\Sigma^+,\pi^0;\Lambda,\Sigma^0)
&=\int\frac{d^4l}{(2\pi)^4}\mathcal{F}(l^2,m_{\Lambda}^2)\mathcal{F}(l^2,m_{\Sigma}^2)\bar u_\psi(q)i\left(\hat CP_{\mathcal R}\right)\frac{i(\slashed q+m_{\Sigma^0})}{q^2-m_{\Sigma^0}^2 }i\left( g_{\Sigma^0\Lambda\pi^0}\gamma_5\right)
\nonumber\\
&&\frac{i(\slashed l+m_{\Lambda})}{l^2-m_{\Lambda}^2 }i(g_{\Sigma^+\Lambda\pi^+}\gamma_5)\frac{i(\slashed l+\slashed p+m_{\Sigma^+})}{(l+p)^2-m_{\Sigma^+}^2 }i( g_{\Sigma^+\Sigma^+\pi^0}\gamma_5)\notag\\
&&\frac{i}{(q-l)^2-m_{\pi^0}^2 }u_{\Sigma^+}(p+q),
\nonumber\\
&\mathcal{T}_a^{\prime}(\Lambda,\Lambda)&= \bar{u}_\psi(q)i(\hat CP_{\mathcal R})\frac{i(\slashed q+m_{\Lambda})}{q^2-m_{\Lambda}^2}(-\frac{\kappa_{\Lambda}}{2m_{\Lambda}}\sigma^{\mu\nu}q_{\nu})u_{\Lambda}(p+q)\epsilon_{\mu}^*(q),\notag\\
&\mathcal{M}^{\prime}_b(\Sigma^+,\pi^-;\pi^+)
&=\int\frac{d^4l}{(2\pi)^4}\mathcal{F}(l^2,m_{\pi}^2)\bar u_\psi(q)i\left(-i\frac{\alpha\hat C}{2f}P_{\mathcal R}\right)\frac{i(\slashed q+\slashed l+m_{\Sigma^+})}{(q+l)^2-m_{\Sigma^+}^2 }i(g_{\Lambda\Sigma^+\pi^-}\gamma_5)
\nonumber\\
&&u_{\Lambda}(p+q)\frac{i}{(p-l)^2-m_{\pi^-}^2 }i\left[ e\,g_{\gamma\pi^-\pi^+}(p-2l)^\mu\right]\frac{i}{l^2-m_{\pi^+}^2 }\epsilon_\mu^*(p),
\nonumber\\
&\mathcal{M}^{\prime}_b(\pi^+,\Sigma^-;\Sigma^-)
&=\int\frac{d^4l}{(2\pi)^4}\mathcal{F}(l^2,m_{\Sigma}^2)\bar u_\psi(q)i\left(-i\frac{\alpha\hat C}{2f}P_{\mathcal R}\right)\frac{i(\slashed l+m_{\Sigma^-})}{l^2-m_{\Sigma^-}^2 }
\nonumber\\
&&\left[-ie\left(Q_{\Sigma^-}\gamma^\mu-i\frac{\kappa_{\Sigma^-}}{2m_N}\sigma^{\mu\nu}p_\nu\right)\right]
\frac{i(\slashed p+\slashed l+m_{\Sigma^-})}{(p+l)^2-m_{\Sigma^-}^2 }i(g_{\Lambda\Sigma^-\pi^+}\gamma_5)u_{\Lambda}(p+q)
\nonumber\\
&&
\frac{i}{(q-l)^2-m_{\pi^+}^2 }
\epsilon_\mu^*(p),
\nonumber\\[2mm]
&\mathcal{M}^{\prime}_b(\Sigma^+,\rho^-;\pi^+)
&=\int\frac{d^4l}{(2\pi)^4}\mathcal{F}(l^2,m_{\pi}^2)
\bar u_\psi(q)
i\left(-i\frac{\alpha\hat C}{2f}P_{\mathcal R}\right)
\frac{i(\slashed q+\slashed l+m_{\Sigma^+})}{(q+l)^2-m_{\Sigma^+}^2 }
\nonumber\\
&&\left[f_{1,\Lambda\Sigma^+\rho^-}\gamma^\nu-i\frac{f_{2,\Lambda\Sigma^+\rho^-}}{m_\Lambda+m_{\Sigma^+}}\sigma^{\nu\lambda}(p-l)_\lambda\right]\frac{i\left(-g_{\nu\rho}+\dfrac{(p-l)_\nu(p-l)_\rho}{m_{\rho^-}^2}\right)}{(p-l)^2-m_{\rho^-}^2 }
\nonumber\\
&&\left[i\frac{e}{m_{\pi^+}}g_{\gamma\pi^+\rho^-}\epsilon_{\mu\alpha\rho\beta}p^\beta l^\alpha\right]\frac{i}{l^2-m_{\pi^+}^2 }\epsilon^{*\mu}(p)u_{\Lambda}(p+q).
\end{eqnarray}

\end{document}